\documentclass[final]{IEEEtran}

\usepackage{cite}
\usepackage{amsmath}
\usepackage[latin1]{inputenc}
\usepackage{graphicx}

\begin{document}

\title{The Dynamics of A Self-Forming Network}

\author{Igor~Sobrado and~Dave~Uhring%
\thanks{I. Sobrado and D. Uhring are with Forté Computer Systems, Inc.,
  110 East Main Street, Collinsville, Illinois 62234, USA.}}

\maketitle

\begin{abstract}
This article describes our strategy for deploying self-forming ad hoc
networks based on the Internet Protocol version 6 and evaluates the
dynamics of this proposal.  Among others, we suggest a technique
called adaptive routing that provides secure intelligent routing
capabilities to computer communication networks.  This technique uses
the flow label, supports hybrid metrics, network load sharing, and is
not restricted to evaluation of performance on first hop routers when
making routing decisions.  Selective anycasting is an extension to the
anycast addressing model that supports exclusion of members of groups
that perform poorly or inappropriately on a per-host basis.
Distributed name lookup is suggested for integrating self-forming and
global networks where they coexist.  At last, we pose an address
hierarchy to support unmanaged discovery of services in unknown
networks.
\end{abstract}


\section{Introduction}
\PARstart{S}{elf-forming} \emph{ad hoc} computer networks \cite{darpa:plan}
will become an active research field in the coming years.  As other
self-organizing networks, these networks are able to respond to
hostile actions such as \emph{Denial of Service} (DoS) and
\emph{Distributed Denial of Service} (DDoS) attacks more efficiently
than traditional networks.  This ability is useful for deploying
unmanaged computer networks.  Self-forming networks are an adequate
platform to deploy proposals like intelligent autonomous agents
\cite{sobrado:agents} that require some degree of survivability in the
network infrastructure.

A fault tolerant network like the one suggested above requires
intelligent routing capabilities and a technique for discovering and
allocating resources in a unmanaged and non-centralized way.
Requirements include:
\begin{itemize}
\item Reliable, fault tolerant, communication networks supporting an
  intelligent routing framework and redundancy;
\item Discovery of devices offering services in a dynamic networking
  environment, in an unmanaged way;
\item Integration with existing network infrastructures where
  available, supporting a world-wide reaching technique;
\item Automatic configuration of devices; and, finally,
\item A secure network infrastructure.
\end{itemize}

In this paper we propose a technique, called \emph{adaptive routing},
that provides secure intelligent routing capabilities to computer
networks at an \emph{autonomous system} (AS) level.  This technique,
based on the use of the flow label field, resolves the security issues
associated with other routing proposals in a simple and elegant way.
\emph{Selective anycasting} increases the robustness of anycast
addressing, enabling hosts to selectively reject those members of
anycast groups that do not fit their requirements but are still alive.

The most important contributions of this manuscript are the
development of a secure intelligent routing infrastructure for
computer communication networks, and an extension to anycasting that
significantly increases the robustness and reliability of this
addressing model.  Discovery of services and a distributed name lookup
mechanism, presented initially in \cite{sobrado:autoconf} for the
automatic configuration of IPv6 devices, is applied to self-forming
networks.

The remainder of the paper is organized as follows.  In
Section~\ref{work} we introduce related work.  Section~\ref{notation}
describes the notational conventions used in this article.
Section~\ref{adaptive} outlines our proposal for deploying
self-forming ad hoc networks at a theoretical level.
Section~\ref{evaluation} provides a performance evaluation of our
prototype when compared with current fixed networks.
Section~\ref{security} presents the security weaknesses commonly found
on ad hoc networks and, more specifically, self-forming networks, and
how our proposal manages those security issues.  Some possible
research lines are shown in Section~\ref{future}.  Finally,
conclusions are outlined in Section~\ref{conclusion}.

\section{Related Work}
\label{work}
The \emph{Internet Protocol version 6} (IPv6)
\cite{rfc2463,huitema:ipv6} is a good foundation for deploying
self-forming computer networks.  This communication protocol provides
hierarchical addresses and is a key element for supporting safe
intelligent routing using the flow label field.  This section provides
an overview of research efforts related with our proposal.
\begin{itemize}
\item The \emph{Dynamic Host Configuration Protocol} (DHCPv6)
  \cite{rfc3315} allows passing configuration parameters such as
  network addresses, netmasks, and hostnames to network nodes from a
  DHCP server.
\item The \emph{flow label} field \cite{rfc3697} enables
  classification of packets belonging to a specific stream by the
  $\langle$\emph{label}, \emph{src}, \emph{dst}$\rangle$ triplet.
  This field can be used by the packet classifier in a router to
  efficiently forward traffic for a particular data stream.  As
  routers do not need to parse the option headers, packets can be
  processed faster, increasing effective routers throughput.
\item\emph{Intelligent route controllers} \cite{borthick:multi-homing,
  passmore:optimizers, kerravala:routing} are appliances that make
  routing decisions for multi-homed connections implementing route
  changes in Border Gateway Protocol (BGP) \cite{rfc1771} routers.
  Currently, non-BGP routing is a cost effective solution for networks
  that do not want to run a routing protocol as complex as BGP.  An
  intelligent route controller optimizes traffic routed from a subset
  of the Internet address space to a set of non-overlapping regions
  called clusters.
\item The \emph{Internet Control Message Protocol} (ICMPv6)
  \cite{rfc2463} \textsc{redirect} messages are used by routers to
  inform other nodes of a better first hop toward a destination.
  Considered harmful by security concerned sites, \textsc{redirect}
  messages are not honored by most routers.
\item The \emph{routing extension header} \cite{rfc2460} is an IPv6
  header option used to route packets, either strictly or loosely,
  from a source to a destination host.  It is assumed that, as the
  ICMPv6 \textsc{redirect} messages, the routing header is a security
  concern as a consequence of a lack of an authentication mechanism.
\end{itemize}

As a difference with intelligent route controllers, we propose making
intra-AS routing decisions.  Our proposal is intended to complement,
not to replace, intelligent route controllers.  From all the above, we
conclude that both the ICMPv6 \textsc{redirect} messages and the
routing extension header are not adequate mechanisms to achieve
intelligent routing.  Instead, we suggest using a secure mechanism to
modify the interface to which a flow label is assigned.

\section{Notational Conventions}
\label{notation}
Let us define the set of intermediate systems in the $i$-th route
discovered by the route servers as ${\cal
R}^i=\{r_1^i,r_2^i,\ldots,r_{n_i}^i\}$, where $r_j^i$ is the $j$-th
intermediate system in this route.  Routes are calculated to minimize
the cost, $w^j$, for a set of parameters ${\cal
P}=(p_0;p_1,p_2,\ldots,p_n)$.  In this paper $r_i\rightarrow r_{i+1}$
denotes a link between intermediate systems $r_i$ and $r_{i+1}$.  This
link is not bidirectional; in other words, $r_{i+1}\rightarrow r_i$ is
a different link in our simulation.  We pose the notation
$r_i\leftrightarrow r_{i+1}$ to denote both links simultaneously.  A
brief outline of notational conventions used in this manuscript is
provided in Table~\ref{table_conventions}.

\begin{table}[tb]
\centering
\caption{Notational Conventions Used in this Paper}
\label{table_conventions}
\begin{IEEEeqnarraybox}[\IEEEeqnarraystrutmode\IEEEeqnarraystrutsizeadd{2pt}{0pt}]{x/s/s/x}
\IEEEeqnarraydblrulerowcut\\ &\hfill Symbol\hfill&\hfill
Definition\hfill&\\ \IEEEeqnarrayrulerow\\ &${\cal
P}=(p_0;p_1,p_2,\ldots,p_n)$&set of parameters that define the&\\
&&\hspace{0.075in}requirements\rlap{\textsuperscript{*}} of a packet
stream;&\\ &&\hspace{0.075in}$p_i$ is the weight of the $i$-th
parameter&\\ &${\cal R}^i=\{r_1^i,r_2^i,\ldots,r_{n_i}^i\}$&$i$-th
route discovered in the ad hoc network;&\\ &&\hspace{0.075in}$r_j^i$
is the $j$-th intermediate system in the&\\ &&\hspace{0.075in}route;
$n_i$ is the number of intermediate&\\ &&\hspace{0.075in}systems in
that route&\\ &$r_i \rightarrow r_j$&link between intermediate systems
$r_i$ and $r_j$&\\ &$w^i(p_0;p_1,p_2,\ldots,p_n)$&total cost of the
$i$-th route&\\ &$w_j^i(p_1,p_2,\ldots,p_n)$&cost of the $j$-th
intermediate system, $r_j^i$&\\
&$w_\mathrm{opt}(p_0;p_1,p_2,\ldots,p_n)$&lower cost found&\\
\IEEEeqnarraydblrulerowcut\\
&\IEEEeqnarraymulticol{3}{s}{\scriptsize\textsuperscript{*}For
example: bandwidth, latency, number of hops...}%
\end{IEEEeqnarraybox}
\end{table}

\section{Adaptive Computer Networks}
\label{adaptive}
One of the goals of a self-forming ad hoc computer network is being
able to response to a changing environment (e.~g., degrading softly
under a DoS attack).  Both automatic discovery of services and
adaptive routing are powerful tools for responding to the challenges
introduced by dynamic network topologies.  The former is based on the
use of reliable anycast groups and service oriented IPv6 addresses;
the latter on \emph{route servers} (RSes) and flow labels.  We suggest
using a distributed name service for integration between self-forming
and fixed networks.  This naming service allows nomadic networks to be
reachable without using tunnels.  The use of a \emph{local namespace}
on each device for allocating services discovered simplifies
application management.

\subsection{Discovery of Services}
As outlined in \cite{sobrado:autoconf}, anycasting
\cite{rfc3513,rfc1546} with service oriented IPv6
addresses\footnote{Where the host portion of the IPv6 address has been
replaced by a service identifier field.} can be used to build a
framework for the automatic discovery of machines offering services.
The unicast addresses of those machines can be added to local
namespaces in each self-configurable device to simplify configuration
of applications.  Selective anycasting, described below, can greatly
improve reliability of anycast addressing.

\subsection{Overlay Networks}
\emph{Distributed name lookup} (DNL) \cite{sobrado:autoconf} is a name
resolution technique useful for reaching nodes of a self-forming
nomadic network where access to a global communication infrastructure
is possible.  DNL splits name resolution in two tasks that will run on
probably different nameservers.  In fact, DNL makes forward resolution
in the base network (i.~e., the network of the mobility provider) and
reverse translation in the network where the mobile devices reside.
These temporary resource records cannot be transferred to slave
nameservers.

\subsection{Adaptive Routing}
We suggest using RSes, supporting hybrid metrics for route
optimization, and an intelligent routing based on the flow label
field.  Hybrid metrics allow routing infrastructure to assign a cost
to each intermediate system that depends on more than one parameter.
Each parameter can have a different weight in the estimation of the
cost.

\subsubsection{Combining Multiple Metrics in a Single (Hybrid) Metric}
RSes can assign a cost to each intermediate system as a function of
the requirements for packet forwarding for a given data stream (e.~g.,
high bandwidth, low latency, $\ldots$).  Let us define the total cost
$w^j(p_0;p_1,p_2,\ldots,p_n)$ for a route ${\cal
R}^j=\{r_1^j,r_2^j,\ldots,r_{n_j}^j\}$ as:
\begin{equation}
\label{eqn_cost2}
w^j(p_0;p_1,p_2,\ldots,p_n)\stackrel{\mathrm{def}}{=}\frac{p_0}{b^j}
  +\sum\limits_{i=1}^{n_j} w_i^j(p_1,p_2,\ldots,p_n)\enspace ,
\end{equation}
where ${\cal P}=(p_0;p_1,p_2,\ldots,p_n)$ is a set of parameters that
define the requirements of the hybrid metric; in this equation,
\begin{equation}
\label{eqn_bandwidth}
b^j=\!\!\!\min\limits_{1\le i\le n_j}\!\!\!b_i^j
\end{equation}
is the end-to-end effective bandwidth between the source and
destination hosts ($b_i^j$ is the available bandwidth in the
intermediate system $r_i^j$); $w_i^j(p_1,p_2,\ldots,p_n)$ is the
cost\footnote{$p_0$, the weight assigned to the bandwidth requirement,
must be applied to the effective bandwidth for the end-to-end route.
This parameter cannot be applied to the throughput on each
intermediate system.} of $r_i^j$ in the route ${\cal R}^j$ for ${\cal
P}$.  The best path discovered is the one that minimizes the
end-to-end cost:
\begin{equation}
\label{eqn_optimal}
w_\mathrm{opt}(p_0;p_1,p_2,\ldots,p_n)=\min\limits_{\forall j}
  w^j(p_0;p_1,p_2,\ldots,p_n)\enspace .
\end{equation}
Table~\ref{table_metrics} shows a subset of end-to-end metrics that
can be used to calculate the cost of a route between two hosts.
\begin{table}[tb]
\centering
\caption{Metrics for End-to-End Performance Estimation}
\label{table_metrics}
\begin{IEEEeqnarraybox}[\IEEEeqnarraystrutmode\IEEEeqnarraystrutsizeadd{2pt}{0pt}]{x/s/s/s/s/x}
\IEEEeqnarraydblrulerowcut\\ &&&\hfill Mathematical
Expression\hfill&&\\
&\hfill\raisebox{-3pt}[0pt][0pt]{Symbol}\hfill&\hfill\raisebox{-3pt}[0pt][0pt]{Quantity}\hfill&&\hfill\raisebox{-3pt}[0pt][0pt]{Units}\hfill&\IEEEeqnarraystrutsize{0pt}{0pt}\\
&&&\hfill for this metric\hfill&\\ \IEEEeqnarrayrulerow\\
&$b^j$&available bandwidth\rlap{\textsuperscript{a}}&$b^j=\min_{1\le
i\le n_j}b_i^j$&$\mathrm{kB}\cdot\mathrm{s}^{-1}$&\\
&multiple&communication
cost&$\left\{\!\!\!\begin{array}{l}\mbox{price,}\\\mbox{reliability,}\\\mbox{security,}\\\mbox{etc}\ldots\end{array}\right.$&N/A&\\
&$t$&delay\rlap{\textsuperscript{b}}&$t^j=\sum_{i=1}^{n_j}t_i^j$&$\mathrm{s}$&\\
&$\Delta t$&jitter\rlap{\textsuperscript{c}}&$\Delta
t^j=\sum_{i=1}^{n_j} \Delta t_i^j$,&$\mathrm{s}$&\\
&&&\hspace{0.075in}where $\Delta t_i^j=t_i^j-\tilde t_i^j$ is the&\\
&&&\hspace{0.075in}delay variation in $r_i^j$&\\ &$n_i$&number of
hops&N/A&none&\\ \IEEEeqnarraydblrulerowcut\\
&\IEEEeqnarraymulticol{4}{s}{\scriptsize\textsuperscript{a}For file
transfer protocols.}\\*[-3pt]
&\IEEEeqnarraymulticol{4}{s}{\scriptsize\textsuperscript{b}For
interactive applications (e.~g., TELNET).}\\*[-3pt]
&\IEEEeqnarraymulticol{4}{s}{\scriptsize\textsuperscript{c}For
multimedia streams.}%
\end{IEEEeqnarraybox}
\end{table}

\subsubsection{Routing Packets}
Intelligent routing can usually be abused to gain access to networks
whose firewalls are poorly configured.  Therefore, routing decisions
should not be made by untrusted third parties (e.~g., hosts) but from
authenticated devices.  For adaptive routing, we suggest the use of
RSes that will try to discover the route that best fits the set of
requirements ${\cal P}$ for a data stream between two nodes of the
self-forming network.  These devices must be authorized to modify the
interface assigned to a flow label on routers.  Routers supporting
this feature are called \emph{adaptive routers} in this article.
Adaptive routers can monitor their network interfaces looking for
communication failures; if a failure is detected, adaptive routers can
ask an authorized RS for an alternative route to the destination host.
RSes can use a \emph{keep-alive} mechanism to ascertain the
availability of adaptive routers.  An adaptive router that stops
responding to the requests of a RS is an indication of a network
failure too.

\subsubsection{Selective Anycasting}
Let us suppose that one of the members of the anycast group is not
performing as expected.  The members of the self-forming network
should have a chance to reject nodes that are inadequate or deficient.
Existing keep-alive mechanisms cannot detect members that behave
poorly or inappropriately but are still alive.  Our proposal is using
a members \emph{exclusion header} (EH) to provide a list of machines
that should not be contacted\footnote{As anycast addresses are
assigned to routers \cite{rfc3513} to simplify routing, this extension
header does not require support in the members of the groups.}.  To
protect clients of the self-forming network against variations in the
routing path as a consequence of changes in the network topology, we
suggest using the unicast addresses assigned to the members of the
anycast group instead of its relative position in the routing path.
Anycast addresses can be translated to unicast ones, using either
\emph{anycast address mapper} or the \emph{source identification
option} \cite{masafumi:anycasting}.  Each time an entry is added to
the EH, a new data stream must be established; as a consequence, a new
flow label is calculated by the source host.  This header should be
under the control of end-user nodes because:
\begin{itemize}
\item Routers are not designed for network analysis; and,
\item Applications have the ability to decide if a member of an
  anycast group is performing adequately, and should have a chance to
  reject those members that does not.
\end{itemize}
Selective anycasting is a lightweight extension to the anycasting
addressing model that does not introduce overhead in routing if flow
labels are used\footnote{Flow labels are required for adaptive
routing, not for selective anycasting.}.

\begin{figure}[ht]
\centering
\includegraphics{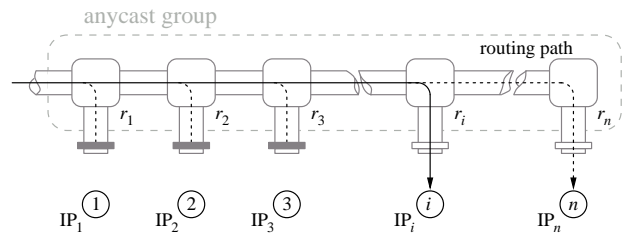}
\caption{A Fluid Mechanics analogy to Selective Anycasting}
\label{fig_selective}
\end{figure}

Fig.~\ref{fig_selective} outlines an analogy between selective
anycasting and a simple mechanical set-up.  Let us suppose that an
incompressible Newtonian fluid flows in a continuous stream on the
pipeline described in this figure.  Joints in this pipeline are
comparable to anycast routers.  Each duct has a valve that acts as a
control device for conveying the Newtonian fluid in the experimental
device.  These valves close temporarily an orifice that permits the
movement of fluid to the ``members of the anycast group'', in the
lower part of the figure.  Initially, all valves are open, allowing
the incompressible fluid to convey to the nearest member of the
anycast group from the point of view of the pipeline topology.  In our
analogy, adding the unicast IP address assigned to a member of the
group to the EH is like closing the valve in the duct that joints that
member to the main pipeline.  Without those valves, the fluid that
flows on the pipeline has no chance to be carried to other members of
the anycast group.  In our scenario nodes whose IP addresses are in
the set ${\cal S}=\{\mathrm{IP}_1, \mathrm{IP}_2, \ldots,
\mathrm{IP}_{i-1}\}$ had been excluded by closing the valves in the
ducts that join them to the pipeline.  These nodes will not be reached
until valves are open again (i.~e., until their addresses are removed
from the EH and a new packet stream to the anycast group is
established).

\section{Experimental Evaluation}
\label{evaluation}
We used the \emph{ns Network Simulator} \cite{breslau:ns, McCanne:ns}
for testing the proposal outlined in this manuscript.  Our prototype
was developed using the \emph{Object Tcl} \cite{wetherall:otcl}
programming language, an extension to the Tool Command Language (Tcl)
\cite{ousterhout:tcl} for dynamic object-oriented programming.  In
this Section, we describe the experimental set-up used to test our
intelligent routing model and provide performance metrics for our
prototype when compared to standard routing proposals.

\subsection{Description of the Prototype}
The aim of our simulation is estimating the ability of our proposal to
recover connectivity when compared to standard routing algorithms
after a part of the network has been damaged; consequently, the
network topology assures the existence of more than one valid route
between the source and destination hosts.  We simulate a damaged link
between intermediate systems $r_i$ and $r_j$ by turning down the links
$r_i\leftrightarrow r_j$ simultaneously.  Same failure conditions are
applied for all routing proposals evaluated in this article.

\begin{figure}[ht]
\centering
\includegraphics{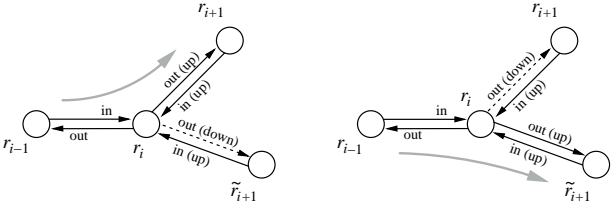}
\caption{Changing the Interface assigned to a Flow Label in our Prototype}
\label{fig_routing}
\end{figure}

Fig.~\ref{fig_routing} shows how we have implemented the flow label
updating mechanism in \emph{ns}.  Let us suppose that an intermediate
system $r_i$ has two output interfaces, $r_i \rightarrow r_{i+1}$ and
$r_i \rightarrow \tilde r_{i+1}$, both of them valid routes toward a
destination host.  To route traffic to one of these interfaces our
prototype turns down all the output links except the one that will
carry the data stream.  In this scenario, both $r_{i+1}\rightarrow
r_i$ and $\tilde r_{i+1}\rightarrow r_i$ remain up to allow
acknowledgments (\textsc{ack}s) reaching the host that has sent the
packets through the link $r_i \rightarrow r_{i-1}$.  Our simulation
uses a \emph{distance vector} (DV) routing algorithm.

\subsection{Performance Evaluation}
The goal for our routing proposal is not performance but reliability.
On the other hand, intelligent routing is a powerful tool for
increasing network performance, allowing routing infrastructure to
make routing decisions based on a global network state, instead of
\emph{first neighbors} feedback.

\begin{figure}[t]
  \begin{minipage}{3.45in}
    \includegraphics[width=3.45in]{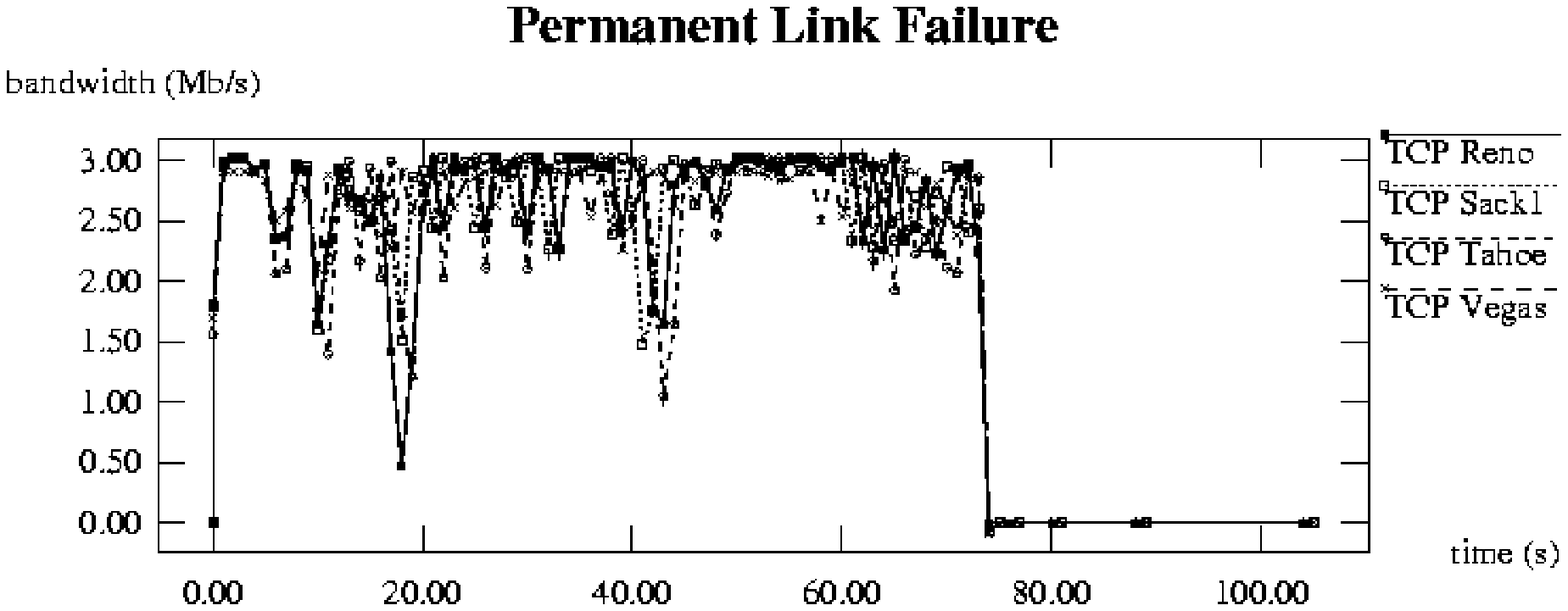}
    \caption{Throughput for Standard TCP in the first Scenario}
    \label{fig_1a}
  \end{minipage}
  \hfill
  \vspace{0.16in}
  \begin{minipage}{3.45in}
    \includegraphics[width=3.45in]{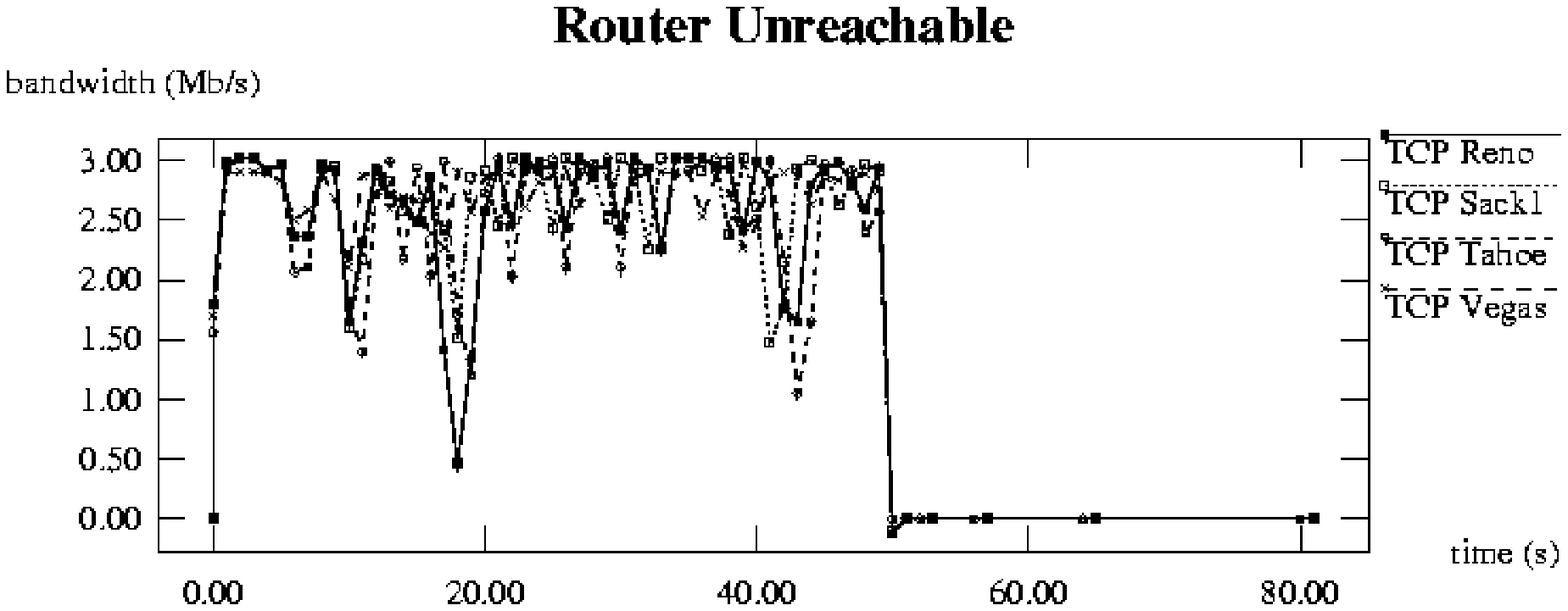}
    \caption{Throughput for Standard TCP in the second Scenario}
    \label{fig_2a}
  \end{minipage}
  \hfill
  \vspace{0.16in}
  \begin{minipage}{3.45in}
    \includegraphics[width=3.45in]{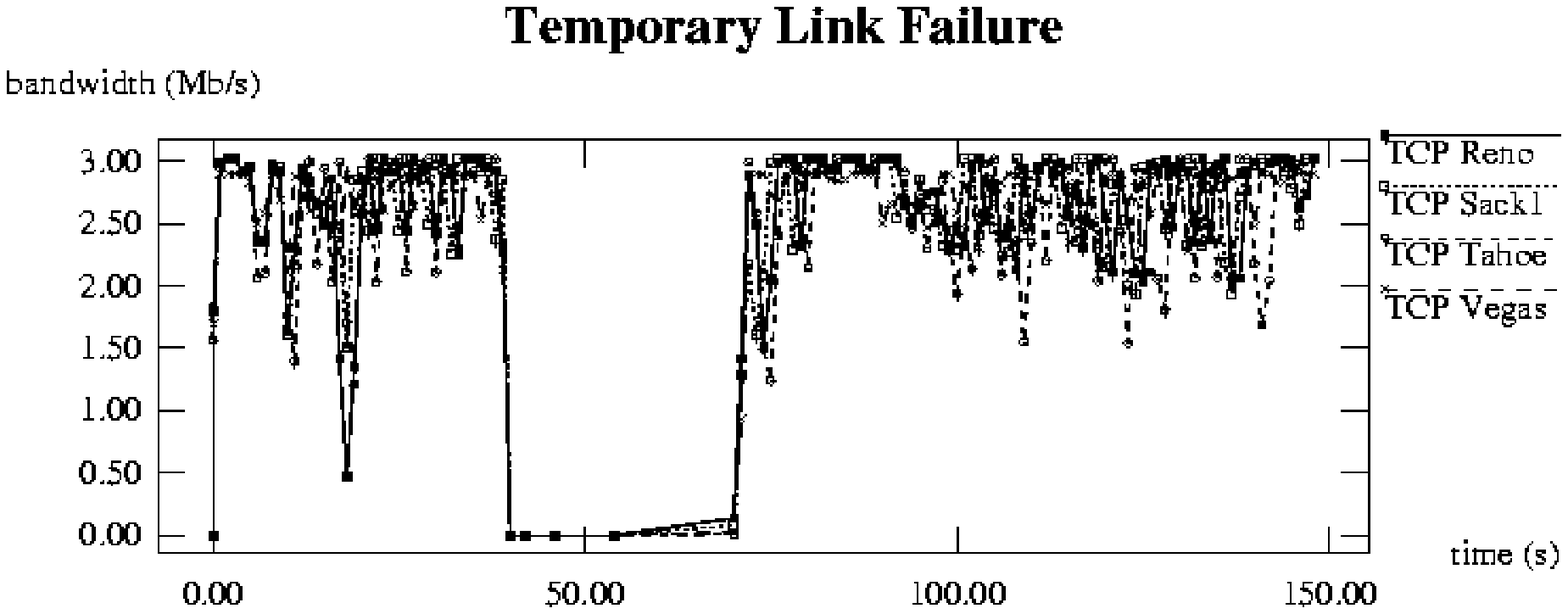}
    \caption{Throughput for Standard TCP in the third Scenario}
    \label{fig_3a}
  \end{minipage}
\end{figure}

\begin{figure}[t]
  \begin{minipage}{3.45in}
    \includegraphics[width=3.45in]{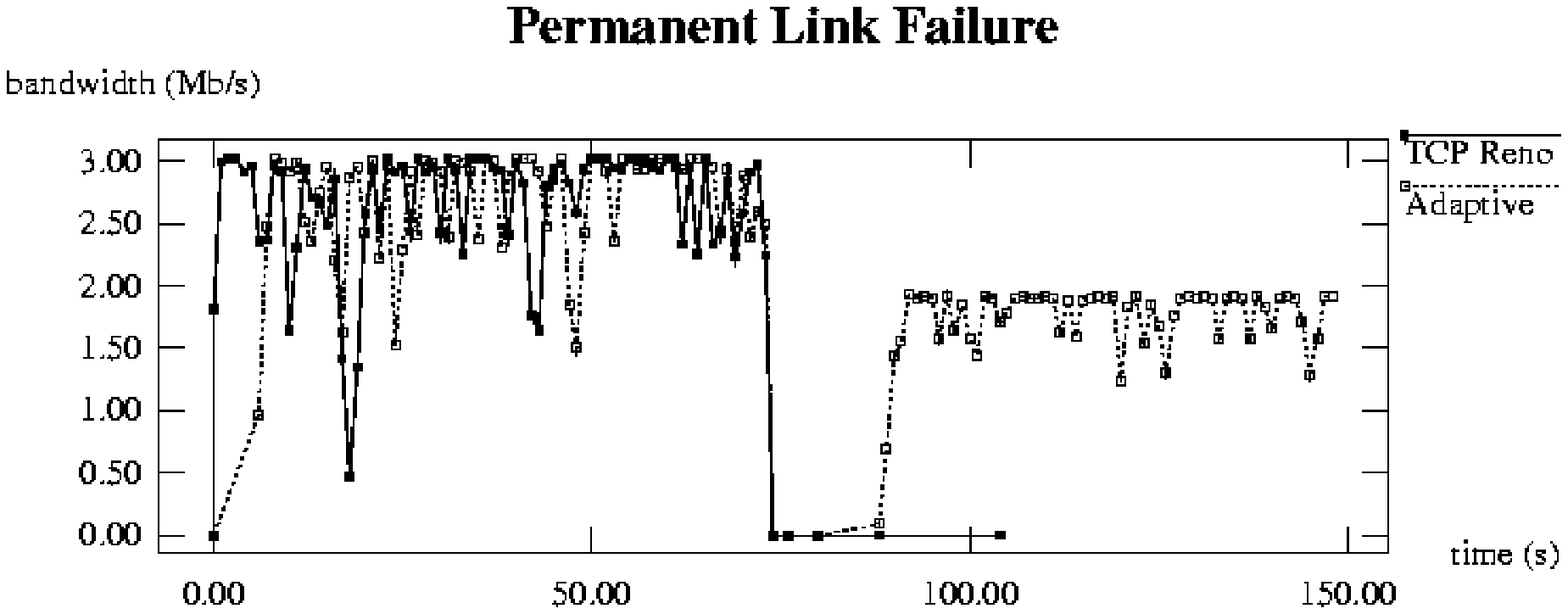}
    \caption{Throughput for the Flows in the first Scenario}
    \label{fig_1b}
  \end{minipage}
  \hfill
  \vspace{0.16in}
  \begin{minipage}{3.45in}
    \includegraphics[width=3.45in]{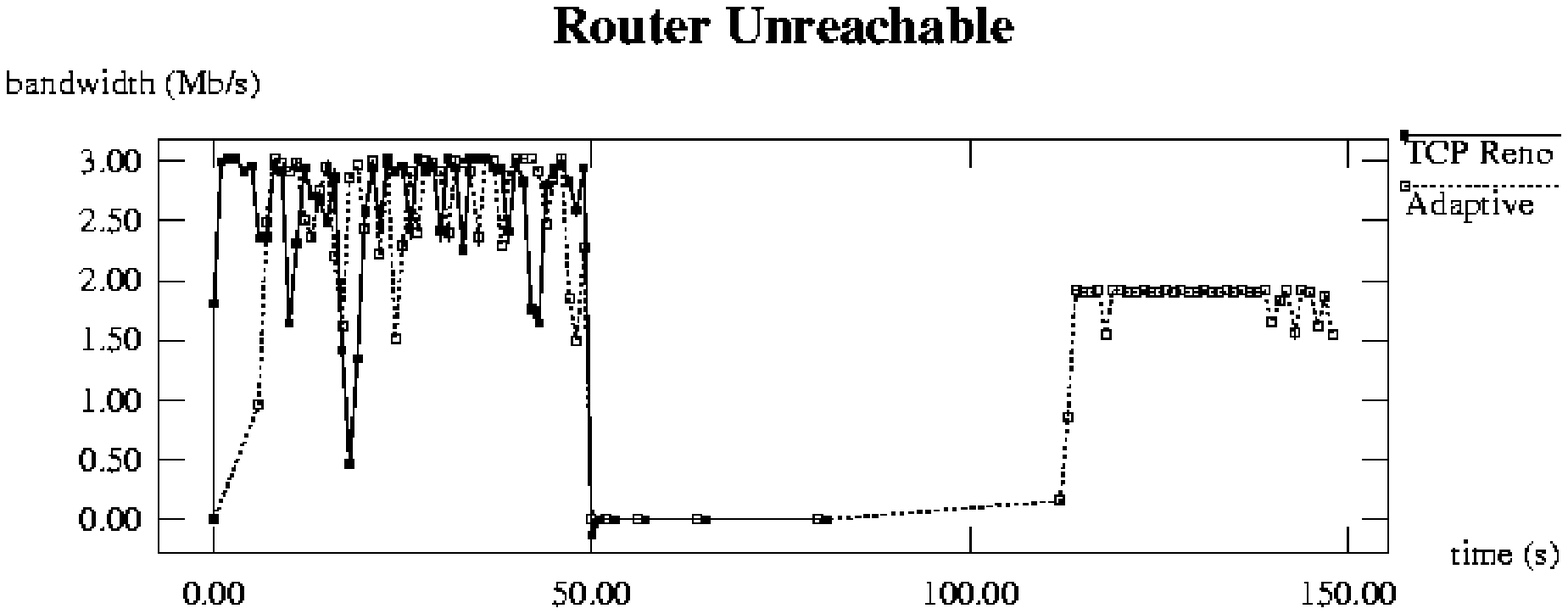}
    \caption{Throughput for the Flows in the second Scenario}
    \label{fig_2b}
  \end{minipage}
  \hfill
  \vspace{0.16in}
  \begin{minipage}{3.45in}
    \includegraphics[width=3.45in]{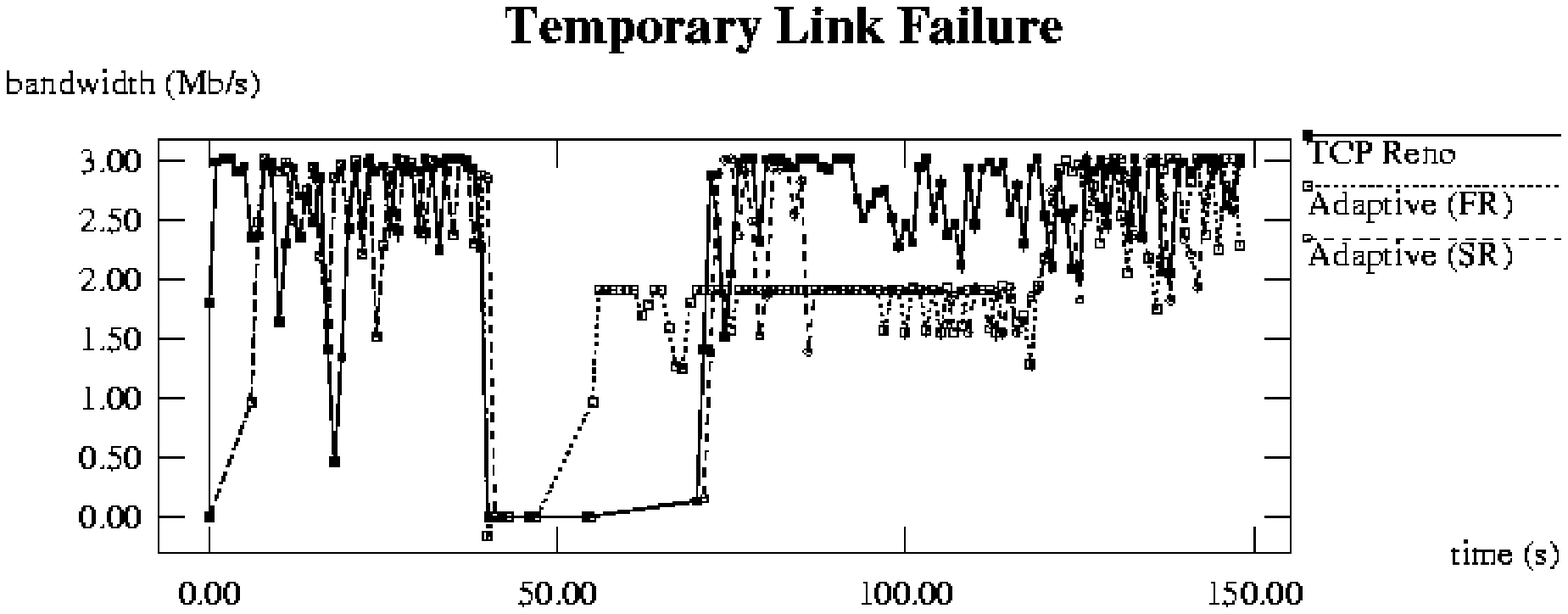}
    \caption{Throughput for the Flows in the third Scenario}
    \label{fig_3b}
  \end{minipage}
\end{figure}

Figs.~\ref{fig_1a} up to \ref{fig_3b} illustrates the performance of
TCP Reno, a \emph{selective acknowledgment} (\textsc{sack}) TCP
sender, TCP Tahoe, TCP Vegas, and our adaptive routing proposal.
Fig.~\ref{fig_1b} depicts network dynamics when a permanent link
failure is detected by an adaptive router and announced to a RS.
Scenario outlined in Fig.~\ref{fig_2b} is a variant of the previous
one; in this case, both a standard router and an adaptive router are
unreachable after the link failure.  A higher delay in recovering
network connectivity in the self-forming network is observed because a
new route is not calculated by the RS before the keep-alive mechanism
ascertains that the adaptive router is not available.  Finally,
Fig.~\ref{fig_3b} depicts the effect of a short loss of connectivity.
When the adaptive route detects the network failure it sends a request
to update the route followed by data streams to a RS.  Both a
\emph{fast response} (FR) from the RS, received before connectivity is
recovered, and a \emph{slow response} (SR), received after recovering
normal network conditions, are compared with the performance of TCP
Reno in same network conditions.

\section{Security Considerations}
\label{security}
Joining anycast and multicast groups in a secure manner
\cite{dondeti:anycast,judge:security} is a requirement for supporting
current networking services.  Authentication of the members of anycast
groups is required for discovery of services.  Selective anycasting
provides reliable and fault tolerant anycast groups.

Adaptive routing works for self-forming networks with an internal
packet forwarding mechanism.  It is a secure approach to intelligent
traffic routing because:
\begin{enumerate}
\item The exact route is not under control of network nodes;
\item Only authorized RSes are able to change the route on the routers
  enabled to support this feature.
\end{enumerate}

As both authentication of RSes and a relatively updated knowledge of
network topology is required, intelligent routing must be done at an
AS level.  Contacting with anycast groups of RSes in other ASes allows
this proposal to be extended to a global computer network like
Internet.

\section{Future Work}
\label{future}
We suggest improving the synchronization mechanism between adaptive
routers and RSes.  Detection of changes in the network topology as
soon as occur is an important goal.  The development of a keep-alive
mechanism between RSes and adaptive routers will contribute to
detection of network failures that isolate adaptive routers from the
rest of the network.

\section{Conclusion}
\label{conclusion}
Survival from failures in communication infrastructure and attacks
against networking equipment requires development of robust, fault
tolerant, computer communication networks.  This article proposes some
techniques to improve reliability of current communication frameworks
and support construction of self-forming ad hoc computer networks.
Our main contributions are:
\begin{itemize}
\item The development of an anycast addressing extension to allow
  applications to reject those members of anycast groups that are not
  performing adequately, but are still alive; and,
\item A framework, based on IPv6 flow labels, that provides
  intelligent routing capabilities to computer networks.
\end{itemize}

Other techniques we have developed in the last years are suggested for
integration with fixed networks and for the unattended configuration
of devices.

\section*{Acknowledgment}
The authors would like to thank Jerry O. Forté, the founder of Forté
Computer Systems Inc., for providing us with the networking
infrastructure required to perform this research.  Lucas Fern\'andez
Seivane helped with an odd problem translating some figures to
Encapsulated PostScript (EPS) format.

\IEEEtriggeratref{99}

\bibliographystyle{IEEEtran}
\bibliography{networks}
\end{document}